\documentclass[cup6b]{cupbook}
\usepackage{graphicx}
\usepackage{natbib}
\title[Rome, Italy, 27--30 April 2009]
      {The coming of age of X-ray polarimetry}
\author{}
\date{}
\begin{document}
\pagenumbering{arabic}


\author[Wu et al.]{Kinwah Wu (University College London) 
  \and Aimee McNamara (University of Sydney) \and Zdenka Kuncic (University of Sydney)}
\chapter{X-ray polarization from accreting white dwarfs and associated systems}

\abstract{ 
  We present our results of Monte-Carlo simulations 
     of polarized Compton X-rays from magnetic cataclysmic variables, 
     with realistic density, temperature and velocity structures in the accretion flow.    
  Our study has shown that the X-ray linear polarization may reach about 8\% 
     for systems with high accretion rates viewed at a high viewing inclination angle.  
  This value is roughly twice the maximum value obtained by previous studies 
     which assumed a cold, static emission region with a uniform density. 
  We also investigate the X-ray polarization properties of ultra-compact double-degenerate binaries 
     for the unipolar-inductor and direct-impact accretor models. 
  Our study has shown negligible X-ray polarization for the unipolar-induction model. 
  However, the direct-impact accretor model may give X-ray polarization 
     levels similar to that predicted for the magnetic cataclysmic variables.  }

\section{Introduction}

 Magnetic cataclysmic variables (mCVs) and Ultra-compact  double degenerate binaries (UCDs) 
   are potential X-ray polarization sources. 
mCVs contain a magnetic white dwarf 
  accreting material from  a low-mass, Roche-lobe filling companion star. 
There are two major types: (i) the AM Herculis binaries (AM Hers, also known as polars) 
  and (ii) the intermediate polars (IPs) 
  (see \citep{Warner95}). 
In AM Hers, the white-dwarf magnetic field ($B \sim 10^7 - 10^8$~G) 
   is strong enough to lock the whole system  into synchronous rotation.  
It also prohibits the formation of an accretion disk,  
   and the  accretion flow is channelled by the magnetic field into the magnetic polar regions of the white dwarf. 
The white dwarf in an IP has a weaker magnetic field ($B\sim 10^6$~G).  
The white-dwarf magnetosphere truncates the inner part of the accretion disk,  
  and the material flow is channelled by the magnetic field from the inner disk rim to the white-dwarf surface. 
For both AM Her and IP, the supersonic accretion flow becomes subsonic abruptly near the white dwarf surface, 
  thereby forming an accretion shock.  
The shock heats up the accreting matter to keV temperature, 
  and X-rays and optical/IR radiation are emitted from the shock-heated gas 
  as it cools and settles onto the white-dwarf surface
  (see e.g.\ \citep{Wu00}).  
  
\begin{figure}
\centering
\vspace*{0.35cm}
\includegraphics[scale=0.3]{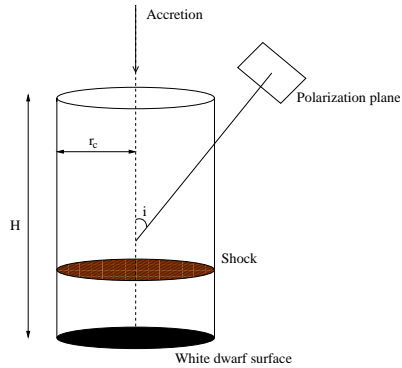}
\caption{
  An illustration of the emission region in mCVs where the polarized X-rays are generated.  
  The accretion column has  a circular cross-section, with a radius $r_{\rm c }$. 
  Compton scatterings occur within a height $H$ from the white-dwarf surface.   } 
\label{scat-column}
\end{figure}
  
A UCD consists of two white dwarfs revolving around each other in a very tight orbit 
  (which has a linear size of Jupiter).  
There are debates on what mechanisms generate the X-rays in UCDs. 
The leading models are the unipolar-inductor (UI) model \citep{Wu02} 
  and the direct-impact accretor (DIA) model \citep{Marsh02}.    
In the UI model, the UCD orbital dynamics are determined jointly 
  by magnetic interaction and gravitational radiation.  
Magnetic induction similar to that in Jupiter and Io sets up an electric current circuit across the binary. 
The electrical dissipation at the foot-points of magnetic field lines connecting the two white dwarfs  
  heats up the white-dwarf atmosphere.   
Because of the convergent field configuration, 
  small hot spots are formed at the surface of the magnetic white dwarf and emit X-rays.  
In the DIA model, the X-rays are accretion powered as in conventional binary X-ray sources. 
The close proximity between the two white dwarfs 
 creates a strong tidal interaction 
 and this prohibit the formation of an accretion disk. 
Mass transfer between the two stars is instead facilitated by a dense material stream.  
X-rays are emitted from a hot spot, where the accretion stream impacts onto the white-dwarf surface. 
The accretion configuration of the DIA model is somewhat similar to that of AM Hers.    

The optical/IR emission from AM Hers is strongly polarized. 
The circular polarization can be as be as high as several ten percents \citep{Wu90}.  
The optical/IR polarization is generated by a thermal cyclotron process \citep{Chanmugam81}, 
  where energetic electrons in the shock-heated plasma gyrate around the magnetic field.  
Optical/IR circular polarization has been observed in several IPs (see \citep{Piirola93},  
  but it is much weaker than that of the AM Hers.  
The origin of the polarization is believed to be the same as that of AM Hers. 
In the UI model, UCDs are predicted to be strong sources of electron-cyclotron masers \citep{Willes04a, Willes04b}. 
The masers are generated through a loss-cone or a shell instability (see \citep{Melrose82})
  developed in the electron population when the charged particles stream along 
  the converging magnetic field lines near the polar regions of the magnetic white dwarf.   
The masers are characterized by a high brightness temperature and almost 100\% circular polarization. 
  
Polarization in the low-energy emission of mCVs is 
   produced by processes involving gyration of electron around a magnetic field. 
Their X-ray polarization is, however, unrelated to magnetic plasma processes. 
Instead it is due to scattering,  
  where unpolarized thermal X-ray photons emitted from the shock-heated region 
  are scattered by the electrons that precipitate onto the white dwarf.     
In this article we show the results of our calculations of X-ray polarization in mCVs 
  and discuss briefly the astrophysical implications of our findings.  
We also show some preliminary results from our study of X-ray polarization 
  in ultra-compact double degenerate binaries.

\begin{figure}
\centering
\vspace*{0.35cm}
\includegraphics[scale=0.32]{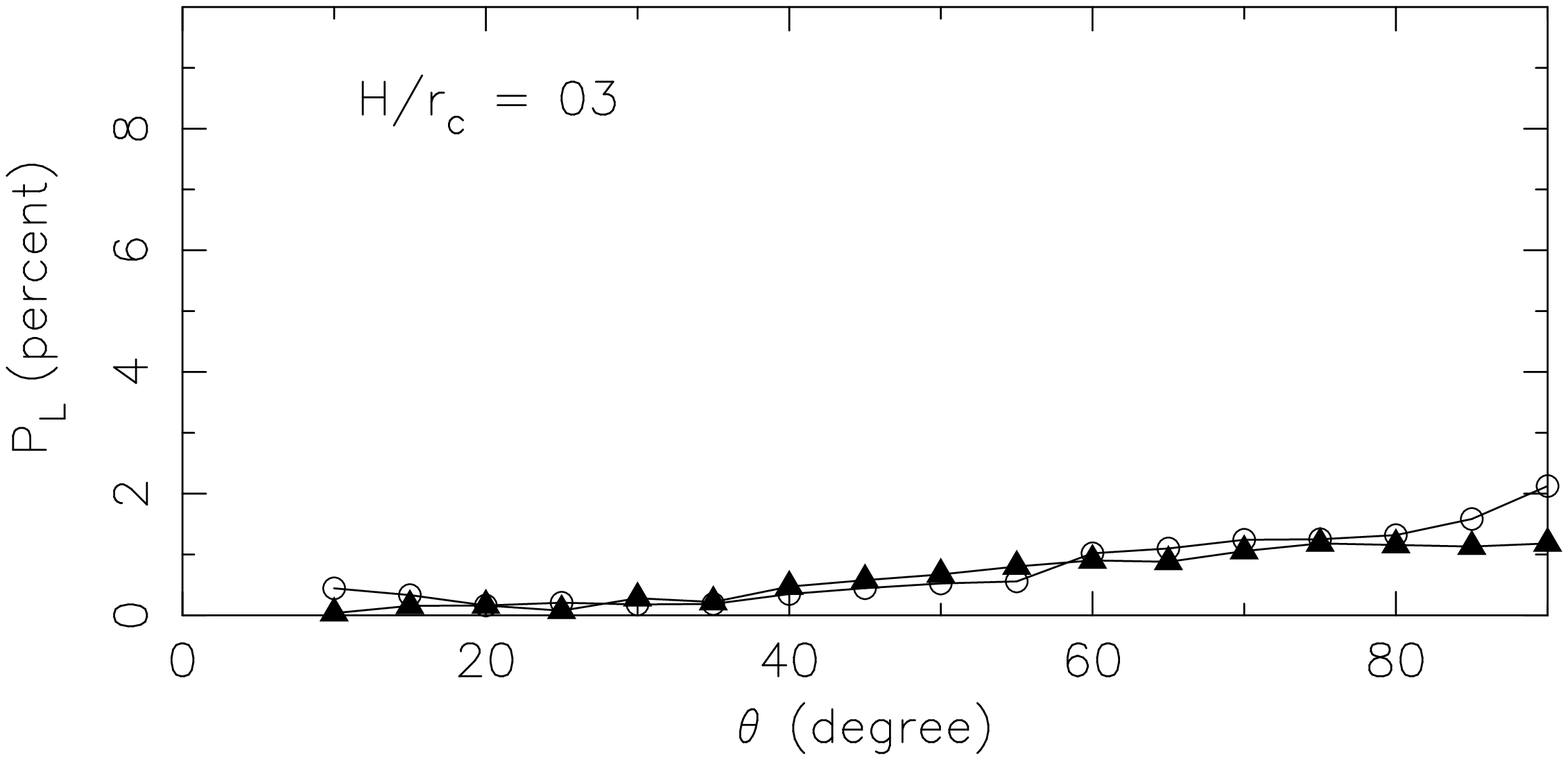} \\ 
\includegraphics[scale=0.32]{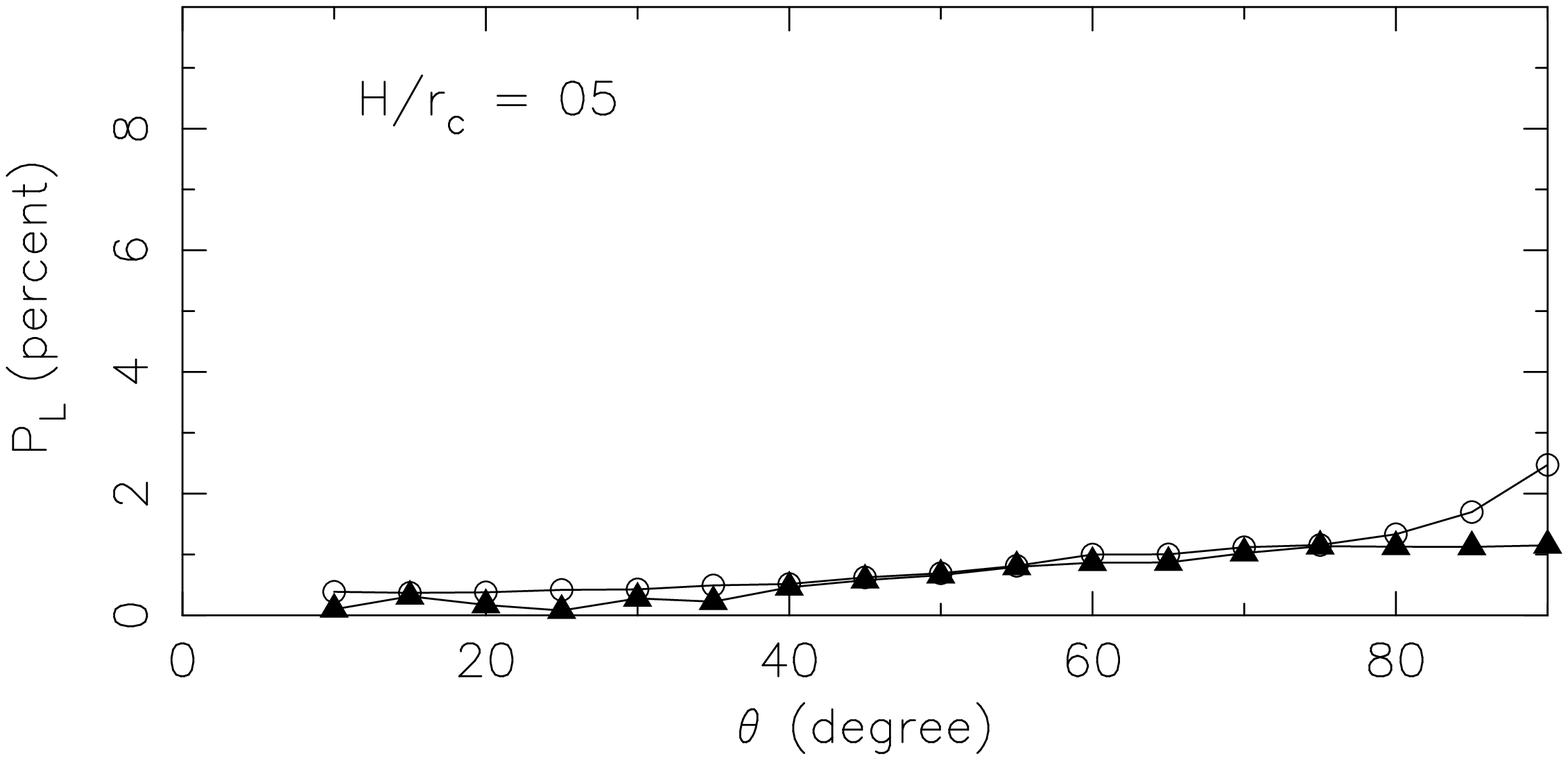} \\ 
\includegraphics[scale=0.32]{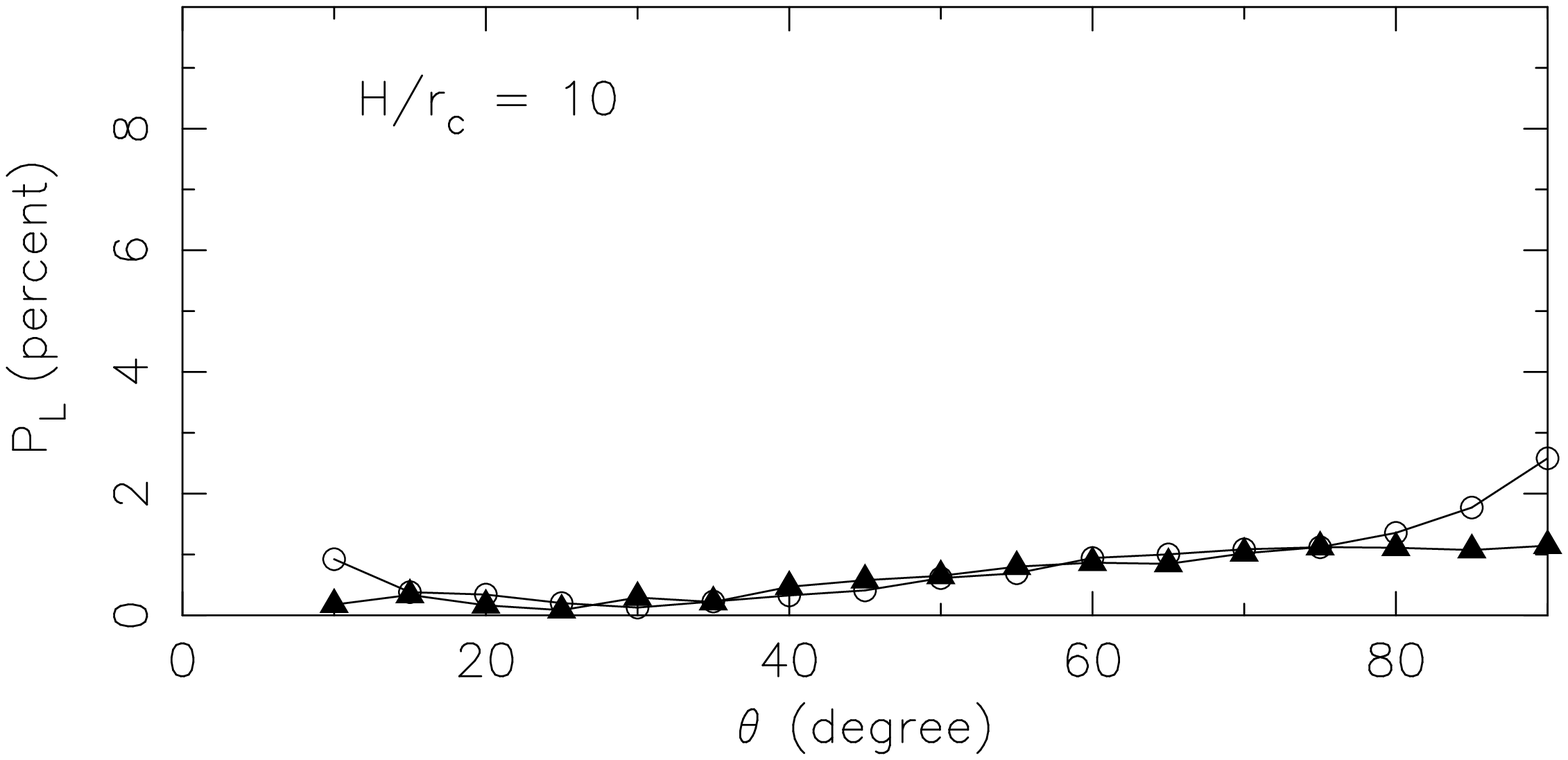}
\caption{
  Linear polarization $P_{\rm L}$ of X-rays as a function of the viewing inclination angle $\theta$ 
     for accretion onto magnetic white dwarfs with masses of 1~M$_\odot$.   
  Filled triangles correspond to the cases with a specific accretion rate 
     ${\dot m} = 1$~g~cm$^{-2}$~s$^{-1}$; and 
    open circles correspond to the case with 10 times higher specific accretion rates. 
  The Thomson scattering optical depths across the accretion shock are $\tau = 0.04$ and 0.3 respectively. 
  The $H/r_{\rm c}$ ratios are 3, 5 and 10 (panels from top to bottom). 
   }
\label{WD10}
\end{figure}

\begin{figure}
\centering
\vspace*{0.35cm}
\includegraphics[scale=0.32]{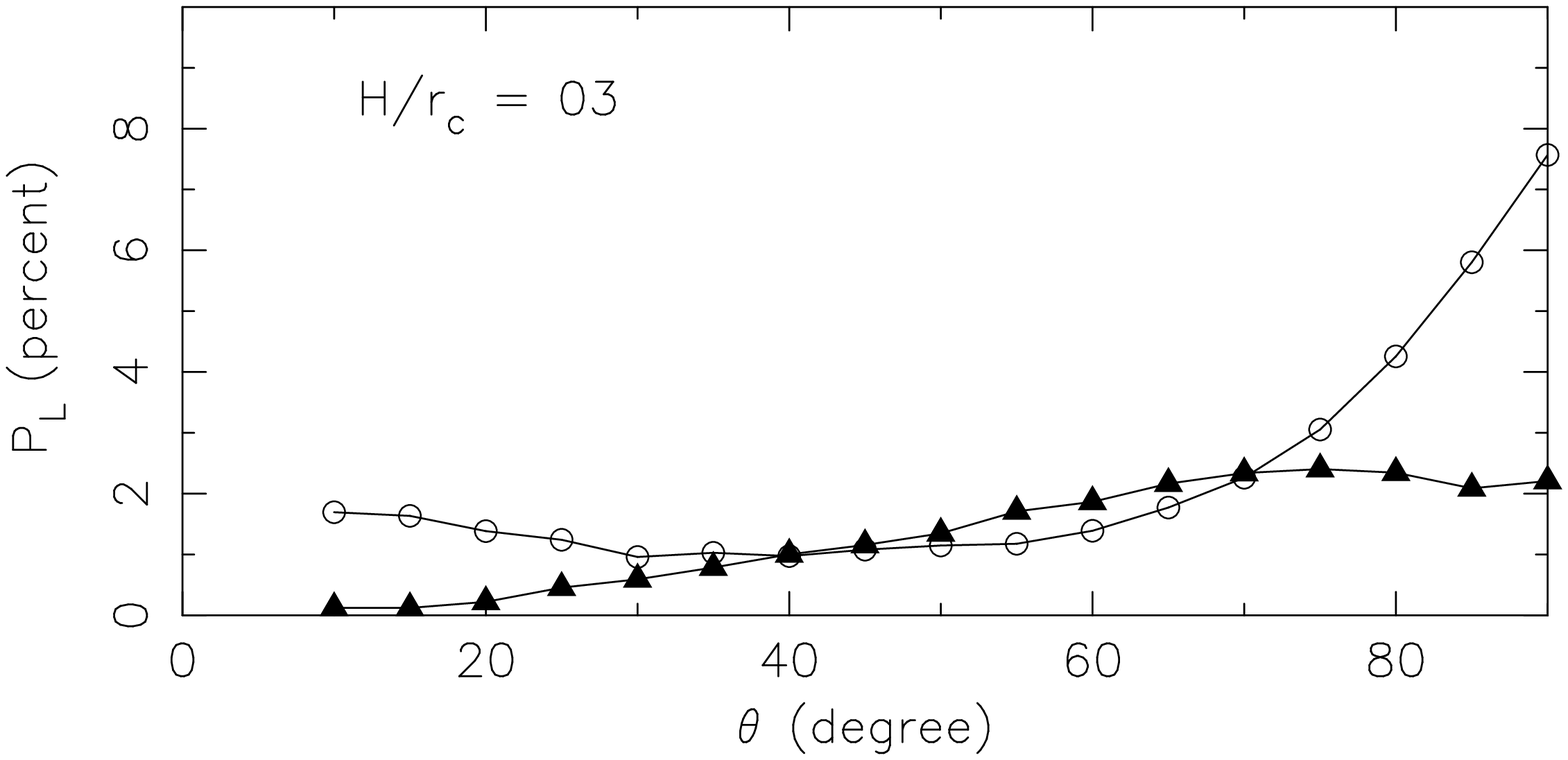} \\ 
\includegraphics[scale=0.32]{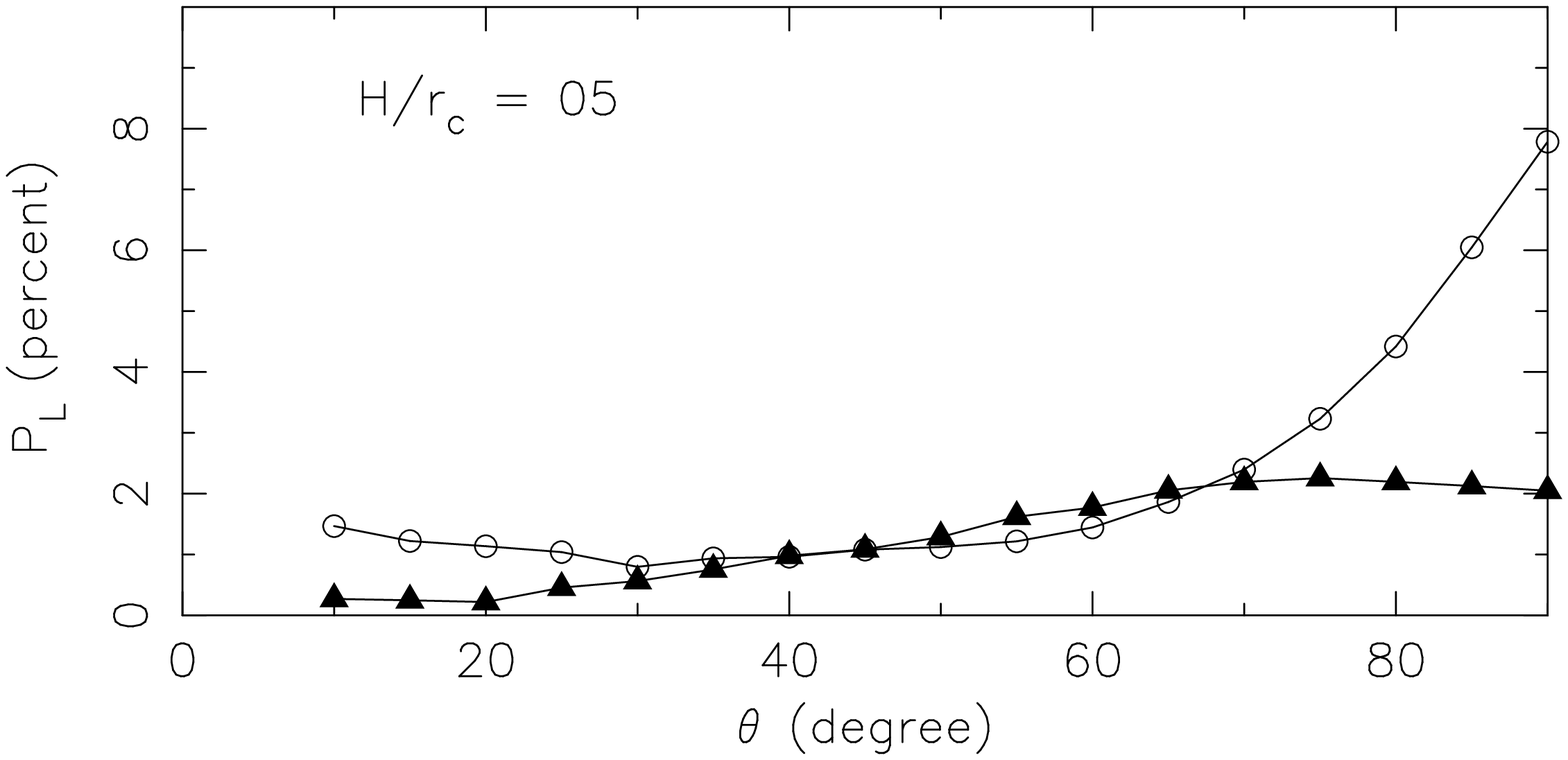} \\ 
\includegraphics[scale=0.32]{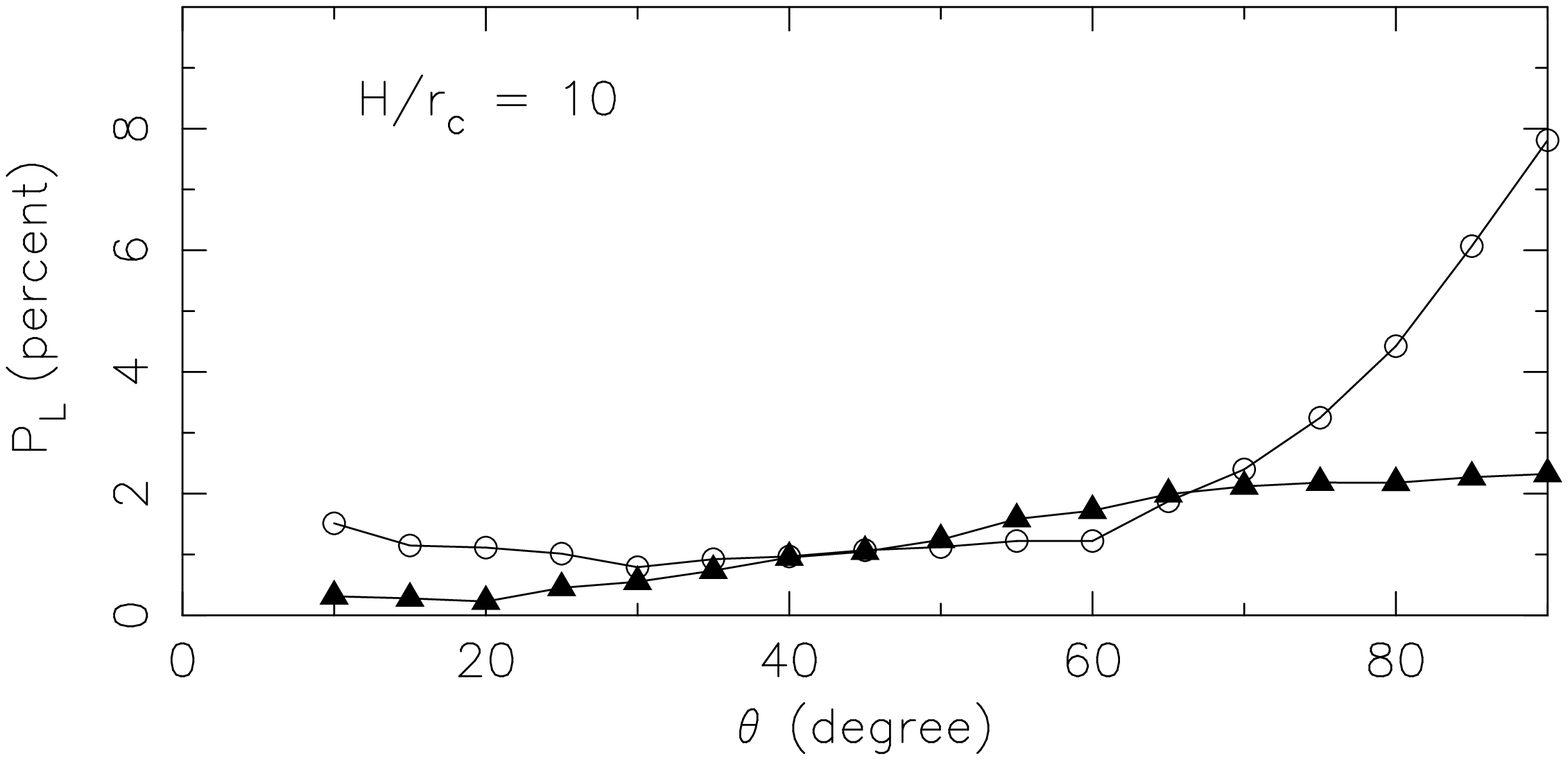}
\caption{Same as Figure~\ref{WD10} for white dwarfs with masses of 0.5~M$_\odot$. 
   The Thomson scattering optical depths across the accretion shock are $\tau = 0.1$ and 1.0 
      for the cases with low and high specific accretion rates respectively. 
      }
\label{WD05}
\end{figure}

\begin{figure}
\centering
\vspace*{0.35cm}
\includegraphics[scale=0.32]{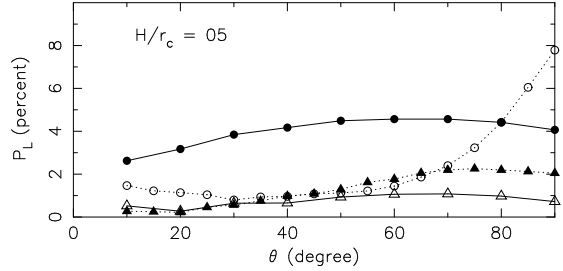} \
\caption{Comparison between the X-ray polarizations  
  predicted by a static, cold emission region with a uniform density (solid lines)
    and by an emission region with structures determined by the hydrodynamic model 
     of \citep{Wu94} (dotted lines). 
  The parameters of the models are the same as those of the case in the middle panel  
     of Figure~\ref{WD05}.  
  Triangles correspond to cases with ${\dot m} = 1$~g~cm$^{-2}$~s$^{-1}$; and 
   circles correspond to cases with ${\dot m} = 10$~g~cm$^{-2}$~s$^{-1}$. }
\label{uniform}
\end{figure}

\begin{figure}
\centering
\vspace*{0.35cm}
\includegraphics[scale=0.32]{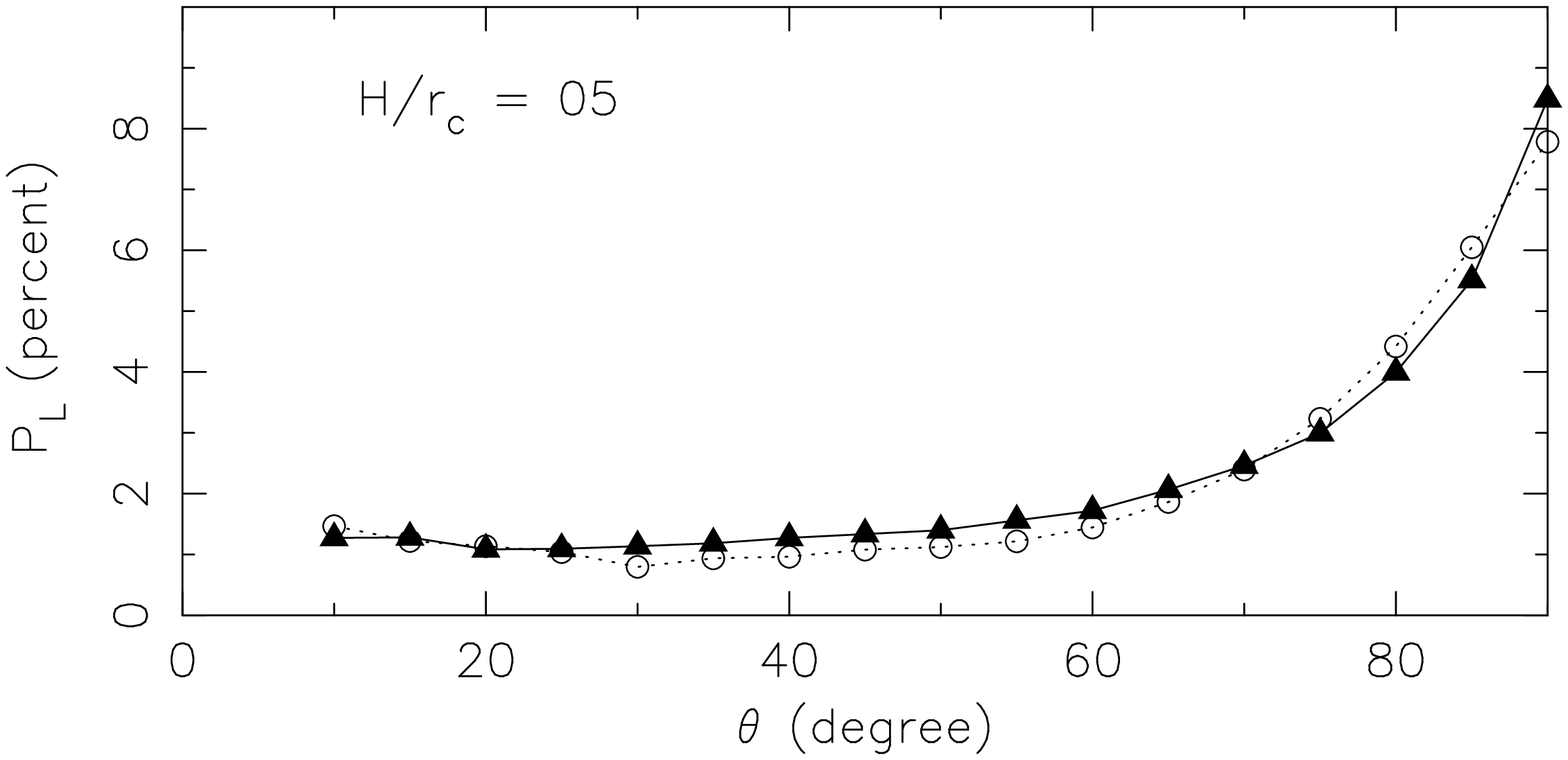} \
\caption{Comparison between X-ray polarizations from mCVs 
    with accretion flows dominated by thermal free-free cooling 
    and by cyclotron cooling. 
  The white-dwarf mass is 0.5~M$_\odot$, 
    and the specific accretion rate is $10$~g~cm$^{-2}$~s$^{-1}$. 
  The ratios of efficiencies of cyclotron to thermal free-free cooling at the shock 
    $\epsilon_{\rm s} = 0$ (open circles) and 10 (filled triangles).  
   }
\label{cyclotron}
\end{figure}

\section{Results and Discussions}

\subsection{X-ray polarization in magnetic cataclysmic variables}  

The geometrical setup in our calculations is shown in Figure~\ref{scat-column}.  
The density, temperature and velocity structure in the post-shock emission region is determined 
   by the hydrodynamics model given in \citep{Wu94, Wu03}.   
Unless otherwise stated, thermal free-free emission is the dominant cooling process. 
We consider a Monte-Carlo approach to simulate the Compton scattering events 
  and the transport of the polarized X-ray photons in the accretion column.  
A non-linear algorithm based on \citep{Cullen01a, Cullen01b} is used 
  to determine the photon mean-free path and the scattering probability. 
The scattering variables are determined following the prescriptions described in  \citep{Pozdnyakov83}.   
The polarizations are calculated using the Klein-Nishina cross section 
  and the formulation for photon-electron scattering given in \citep{Heitler36, Jauch80}. 
The polarized photons that escape from the accretion column are binned 
   according to their energy-momentum 
   and summed to give the spectral polarization 
   at specific intervals of viewing inclination angles. 
The formulation, computational algorithms and numerical simulation procedures 
  are presented in detail in \citep{McNamara08}.  

Figures~\ref{WD10} and \ref{WD05} show the results of two example simulations. 
For the case with a 1.0-M$_\odot$  white dwarf, 
  the maximum value of  the linear polarization $P_{\rm L}$ is about 1$-$2\%. 
For the case with a 0.5-M$_\odot$  white dwarf, 
  the maximum $P_{\rm L}$ may reach about 8\% 
  for viewing angle $\theta \approx 90^{\circ}$. 
The polarization increases slightly with $\theta$ 
  for low specific accretion rates (${\dot m} \sim 1$~g~cm$^{-2}$~s$^{-1}$).     
However, for sufficiently high $\dot m$ ($\sim 10$~g~cm$^{-2}$~s$^{-1}$), 
  the polarization could increase substantially at large $\theta$.  

Figure~\ref{uniform} shows a comparison of the polarization 
  from a structured accretion flow  and a cold, static and uniform density emission region. 
For low accretion rates, the cold, static, uniform density emission region 
  underpredicts the linear polarization, especially at large $\theta$.      
For high accretion rates, it overpredicts the polarization for angles below 
   about $70 - 80^{\circ}$. 
The main difference of the two models  
  is that the structured flow model always has a highly dense base region,  
  where most of the scatterings occur, 
  but the static model has a uniform density throughout the scattering region. 
The underprediction of the polarization by the uniform model at low $\dot m$ 
  is due to the lack of a dense base layer 
  which gives a substantial scattering optical depth. 
The overprediction of the polarization by the uniform model at high $\dot m$  
  is caused by a uniformly high scattering optical depth at all heights above the white-dwarf surface, 
  in contrast to the density drop off with height in the stratified accretion flow.    

\begin{figure}
\centering
\vspace*{0.35cm}
\includegraphics[scale=0.32]{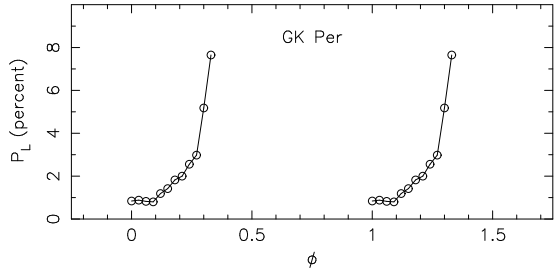} \
\caption{Predicted X-ray linear polarization in the mCV GK Per 
   as a function of the white-dwarf spin phase $\phi$. 
  In the simulations the white-dwarf mass is set to be 0.63~M$_\odot$ \citep{Morales02}, 
     and the specific accretion rate is assumed to be 10~g~cm$^{-2}$~s$^{-1}$. }
\label{GK_Per}
\end{figure}

The high density at the base and rapid density drop off with height are 
  also the reasons why the angle dependence of the polarization is insensitive 
  to the accretion-column aspect ratio in the cases with stratified accretion flows.  
The situation is different for a static, uniform-density scattering region 
  (see \citep{Matt04}), 
  where the effective scattering optical depth depends strongly 
  on the viewing inclination and the aspect ratio of the accretion column.   
Note the polarization is insensitive to the cyclotron cooling process 
  for the same reason, 
  as scatterings occur mainly in the dense base not the less dense region 
  immediately below the shock (Figure~\ref{cyclotron}).  
Figure~\ref{GK_Per} shows the predicted polarization from a system 
  with the same parameters as those derived for the IP GK Per.  
 
\begin{figure}
\centering
\vspace*{0.35cm}
\includegraphics[scale=0.32]{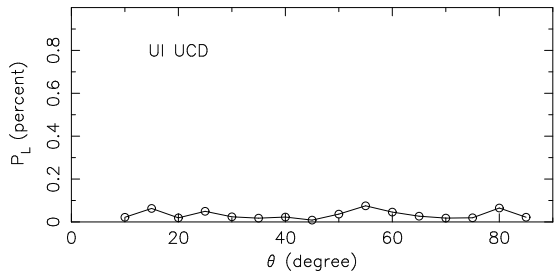} \
\caption{Linear polarization as a function of viewing inclination angle for a UI UCD model. 
   In the model, the bulk Lorentz parameter of the fast streaming electrons $\Gamma = 10^3$, 
      the effective Thomson scattering depth across the column $\tau = 10^{-3}$.  
   The $H/r_c$ ratio is 100.     }
\label{UCD}
\end{figure}

\subsection{X-ray polarization in ultra-compact double degenerates} 

The accretion geometry of the DIA UCD is similar to that of the AM Hers, 
  except that an accretion shock might not be formed. 
In terms of Compton scattering, 
  the model set up would be approximately the same as that of the case 
  with a static, uniform density accretion column. 
Although the gas has a bulk motion in the accretion column, 
  the relatively slow speed implies that 
  Compton recoil is the dominant effect in the scattering event. 
As a first approximation the consequence is not much different from 
  that for the case with cold electrons.   
The polarization is therefore of the order of $1-4$\% (see \citep{Matt04}). 
Thus, as a rough estimate, a polarization of a few percent would be expected 
  from the DIA UCD model. 
The UI UCD does not have a dense accretion column. 
Instead there is a stream of relativistic electrons. 
The dominant effect is Doppler shift instead of recoil. 
Moreover, almost all events are head-on as seen by the relativistic electrons. 
Figure~{UCD} shows our simulation for a model UI UCD. 
As the scattering optical depth is small, there is no substantial polarization. 
X-ray polarization can clearly distinguish the DIA  and the UI model for UCD, 
  despite the fact that the two models could have very similar X-ray spectral and timing properties.

\end{document}